\documentstyle[12pt]{article}
\begin{document}
\thispagestyle{empty}
\begin{flushright}
BI-TP~~93-78
\end{flushright}
\vspace*{15mm}
\font\tenbf=cmbx10
\font\tenrm=cmr10
\font\tenit=cmti10
\font\elevenbf=cmbx10 scaled\magstep 1
\font\elevenrm=cmr10 scaled\magstep 1
\font\elevenit=cmti10 scaled\magstep 1
\newcommand{\Sp}[1]{{\mbox{Li}}_2\left(#1\right)}
\newcommand{\Cl}[1]{{\mbox{Cl}}_2\left(#1\right)}
\newcommand{\rd}{{\rm d}}
\newcommand{\ep}{\varepsilon}
\newcommand{\Pm}{\phantom-}
\newcommand{\Pu}{\phantom1}
\newcommand{\Df}[2]{\mbox{$\frac{#1}{#2}$}}
\newcommand{\Pp}[3]{#1.#2\!\times\!10^{-#3}}
\newcommand{\Fh}[2]{\,{}_#1F_#2}
\newcommand{\Fs}[3]{\!\!\left[\begin{array}{c}#1\,;\\#2\,;\end{array}#3\right]}
\newcommand{\Fu}[2]{\Fs{#1}{#2}{1}}
\newcommand{\Ff}[2]{\Fs{#1}{#2}{4}}
\newcommand{\Fq}[2]{\Fs{#1}{#2}{\frac{1}{4}}}
\newcommand{\Fuq}[2]{\Fs{#1}{#2}{\frac{-q^2}{m^2}}}
\newcommand{\Fum}[2]{\Fs{#1}{#2}{\frac{m^2}{-q^2}}}
\newcommand{\Ffq}[2]{\Fs{#1}{#2}{\frac{-q^2}{4m^2}}}
\newcommand{\Ffm}[2]{\Fs{#1}{#2}{\frac{4m^2}{-q^2}}}
\newcommand{\Ffz}[2]{\Fs{#1}{#2}{z}}
\newcommand{\Fft}[2]{\Fs{#1}{#2}{1-t}}
\newcommand{\Fzz}[2]{\Fs{#1}{#2}{1-z}}

\newcommand{\xip}{\xi^\prime(1)}
\newcommand{\Xp}{X^\prime(0)}
\newcommand{\ei}{(\exp({\rm i}\theta))}
\newcommand{\dfr}[2]{\mbox{$\frac#1#2$}}
\parindent=3pc
\baselineskip=10pt
\begin{center}
              {{\tenbf CALCULATION OF FEYNMAN DIAGRAMS
              \vglue 10pt
               FROM THEIR SMALL MOMENTUM EXPANSION }\\}

\vspace{16mm}

{\large J.~Fleischer,~~~~~~~~~}%
\newcommand{\st}{\fnsymbol{footnote}}%
\medskip
{\large O.~V.~Tarasov $^{1,2}$ }

\medskip
{\em   Fakult\"at f\"ur Physik, Universit\"at Bielefeld
                  D-33615 Bielefeld 1, Germany}

\bigskip

\vspace*{20mm}

\textwidth 120mm
\begin{abstract}

\vglue 0.3cm
{\rightskip=3pc
\leftskip=3pc
\tenrm\baselineskip=12pt
\noindent
 A new  powerful method to calculate Feynman diagrams
is proposed. It consists in setting up a
 Taylor series expansion  in the external momenta squared
(in general multivariable).
The Taylor coefficients are obtained from the original
diagram
by differentiation and putting the
external momenta  equal to zero,
which means a great simplification.  It is demonstrated that it is
possible to obtain by analytic continuation of the
original series high precision numerical values
of the Feynman integrals in the whole cut plane.
For this purpose conformal mapping and subsequent
resummation by means of Pad\'{e} approximants
or Levin transformation are applied.
\vskip 6.0cm
\vglue 0.6cm}

\end{abstract}
\end{center}

\bigskip

\footnoterule\noindent
{\small
$^1$) Supported by Bundesministerium f\"ur Forschung und Technologie.\\
$^2$)On leave of absence from Joint Institute for Nuclear Research, Dubna, 101
000 Moscow,
Russian Federation.}

\textwidth 170mm
\newpage

\setcounter{page}1

\renewcommand{\thefootnote}{\arabic{footnote}}
\setcounter{footnote}{0}

{\elevenbf\noindent 1. Introduction }
\vglue 0.2cm

 Standard-model radiative corrections of high accuracy
have obtained
growing  attention lately in order to cope with the
increasing precision of LEP experiments \cite{LEP}.  In particular
two-loop calculations with  nonzero masses became
relevant \cite{2loop}. While in the one-loop approach
there exists a systematic way of performing these calculations
\cite{1loop}, in the two-loop case there does not exist such
a developed technology and only  a series of partial
results were obtained \cite{methods}, \cite{asy}
but no systematic approach was formulated.
In the present work we demonstrate exactly such an approach.
We also show how to deal with the one-loop case in order
to obtain a possibly faster numerical program
and beyond that we demonstrate that our procedure allows to attack
problems in higher loop calculations which could not be
dealt with by other methods.

   Our approach consists essentially in
performing a Taylor series expansion in terms of external momenta squared
and analytic continuation into the whole  region of kinematical
interest. Simple as
this may sound, there are some unexpected methodical advantages compared
to other procedures.
Taylor expansions of two-point functions have been investigated before
\cite{Recur}, \cite{bft}, but applications were restricted
in particular to the calculation below thresholds.
In  the present paper it is demonstrated  how to properly
use the small momentum expansion such that it can be made
a general, effective and reliable procedure for an unlimited
number of applications.

   Considering a Taylor series expansion in terms of one external momentum
squared, $q^2$ say, the differential operator by the repeated application
of which the Taylor coefficients are obtained, subsequently setting
$q = 0$, is
\begin{equation}
    \Box_q = \frac{{\partial}^2}{\partial{q_{\mu}}\partial{q^{\mu}}}.
\end{equation}

   The extension to the multivariable case is straightforward and will
be discussed for the case of the three-point functions.
Calculating any process from
Feynman diagrams, we can first of all construct scalar amplitudes
{\it before} loop-integrations for the complete process by contracting
with external momenta or taking appropriate traces in the case of
fermion amplitudes. The obtained scalar amplitudes are then integrals
over a combination of scalar propagators with numerators containing scalar
products of integration and external momenta.
In the 2-loop 3-point case, the 3-loop 2-point case, and the 4-loop
vacuum case, e.g. there are 10 scalar products  of 4 momenta, but
only 9 (internal or external) lines against which to cancele these.
 There exist different
approaches of how to deal with these numerators: in the first method
recursion relations are used.
Here one allows for an extra "scalar propagator" (10 th invariant)
with momentum flow different from those in the original diagram.
For the 2-loop bubble diagram, dealt with in the present
paper, this type does not occur.
In the second method  the remaining numerator scalar
products
 are rewritten explicitly as uncontracted products of $d$
- dimensional vectors,
yielding tensor integrals of various types.

   In the one-loop case both methods work and are applied
(see \cite {Davyd} for recursion relations and e.g. \cite {PassV}
for the tensor method). In the two-loop case both methods
work for self-energies (see \cite {bft}, \cite{FJT} for recursion relations and
\cite {Bolle},\cite {WSchB} for the tensor method). For two-loop
vertex functions to our knowledge so far none of these
methods was successfully applied.

   As one can  easily see, our approach works also for these
two-loop vertices and beyond (four-point functions and three-loops):
a differentiation like (1) can be applied to the complete integrand of any
amplitude and only scalar integrals remain with zero
external momenta.
These "bubble diagrams" are essentially the same
(after partial fraction decomposition) for two-point, three-point,
{\ldots} functions for a given number of loops
and we  stress that it is indeed a great technical
simplification
to have to perform Feynman integral calculations only for external
momenta equal zero  even if these integrals contain now arbitrary high powers
 of the scalar
propagators. For their calculation recurrence relations are quite
effective (\cite {Bij/Ve},\cite {Recur}). Beyond
that, it is shown in the present work, that an explicit compact form
of the integrand of the Taylor coefficients can be found. Since higher
powers of
scalar propagators can also be produced by differentiation w.r. to
squared masses, in principle the Taylor coefficients can be obtained by
differentiating a "generating function" (which is the original diagram with
all external momenta equal zero). An explicit
form of this kind is given for a particular one-loop three-point function for
arbitrary masses.

   In several examples it is demonstrated that once the Taylor coefficients
are calculated, our method of analytic continuation yields
miraculously precise
results in the whole kinematical area. Because of this, the final
and most important property of our approach is that the Taylor coefficients
need to be calculated only once for all such that having them (saved
somewhere), the corresponding (class of) Feynman diagrams are (is) done
once forever, the final summation of the series being in any case merely a
relatively small numerical task.

 A final comment appears to be in order concerning the
asymptotic expansion in large external momenta squared,
see  \cite{asy}, \cite{bft}.
Their general property is that they contain logarithmic
 terms, which on the cut yield also the imaginary part properly.
Acceleration methods to improve convergence have also
been applied here \cite{bft}, yet only above the highest
threshold.
It might be possible, however, to perform also for such an expansion
a mapping in the variable $1/q^2$ and thus to penetrate the highest
threshold from above. This has not been tried out so far.
In any case, as our applications demonstrate,
and in particular that one for the electron propagator integral
$I_3$, our method gives also far beyond the highest threshold
(and of course also between thresholds) results
with any desired precision as long as one has calculated
sufficiently many Taylor coefficients. Nevertheless, the two
approaches may be complimentary to each other such that
they can be used in different kinematical domains
and/or for mutual checks.

Our paper is organized as follows:
 in Sect.2 we recall how the Taylor series is set up in
the simplest case of two-point functions and give some general
formulae used for the calculation of their Taylor coefficients.
In Sect.3 we describe the application of our method
to three-point functions. In Sect.4 the method of analytic
continuation is explained and in Sect.5 various applications are
worked out which demonstrate the high precision and
effectivity of the method.  Sect.6 contains our conclusions.

\vglue 0.6cm
{\elevenbf\noindent 2. Expansion of two-point functions in terms
their external momentum squared}

According to the above, the expansion of any two-point function
can be written as (see also Ref. \cite{Recur}):

\begin{equation}
\label{taylor2}
J(\{m^2_i\},q^2)=\sum^{\infty}_{j=0} \frac{1}{j! (d/2)_j}
 \left( \frac{q^2}{4}\right)^j~~ \left( \Box^j_qJ(\{m_i\},q^2)
\right)|_{q=0}
\end{equation}

%\vskip 0.3cm

where

\begin{equation}
(a)_j \equiv \frac{\Gamma(a+j)}{\Gamma(a)}~~~~
{\rm and}~~~~~~d=4-2\varepsilon.
\end{equation}

\vskip 1.0cm

 For one-loop integrals the analytic  result is
well known.
The coefficients of their  expansion in terms of the external
momentum squared by means of partial fraction decomposition
 can be expressed by simple bubble integrals:
\vskip 0.3cm
\begin{equation}
\label{tadpole}
\int \frac{d^d~k_1}{[k_1^2-m_1^2]^{\beta}}=
i\pi^{d/2}~\frac{(-1)^{\beta} \Gamma(\beta-\frac{d}{2})}
{\Gamma(\beta) ~(m_1^2)^{\beta- \frac{d}{2}}} .
\end{equation}

\vskip 0.3cm

The most general scalar  two-loop self-energy integral is
given as a special case or as a combination
of different integrals of the type:

\vskip 0.3cm

\begin{equation}
J(\{m^2_i\},q^2)=
\int \int \frac{d^d~k_1~~d^d~k_2}{D}
\end{equation}

\vskip 0.1cm

with
$$
D=
[k_1^2-m^2_1]^{j_1} [k^2_2-m_2^2]^{j_2} [(k_1-q)^2-m_3^2]^{j_3}
[(k_2-q)^2-m_4^2]^{j_4} [(k_1-k_2)^2-m_5^2]^{j_5} .
$$

The small momentum expansion of these integrals is well
described in \cite{Recur}.
Its Taylor coefficients  can in any case be expressed by
means of a partial fraction decomposition
in terms of bubble integrals
with only three factors in the denominator:
\vskip 0.5cm
\begin{equation}
 J_{\alpha \beta \gamma}=
\int \frac{d^d~k_1~d^d~k_2}{[k_1^2-m_1^2]^{\alpha}~
 [k_2^2-m_2^2]^{\beta} ~[(k_1-k_2)^2-m^2_3]^{\gamma}}.
\end{equation}

 The evaluation of these integrals was first performed in
\cite{Bij/Ve}. In \cite{Recur} it was shown that
the integrals $J_{\alpha \beta \gamma}$
with arbitrary $\alpha, \beta,\gamma$ and arbitrary masses
can be expressed as  a combination of hypergeometric
Appell functions $F_4$ depending on  two dimensionless variables
\begin{equation}
\label{eq:xy}
   x \equiv \frac{m_1^2}{m_3^2}
      \hspace{0.7cm} \mbox{and} \hspace{0.7cm}
   y \equiv \frac{m_2^2}{m_3^2}.
\end{equation}

For practical calculations, however, it is more effective to
use recurrence relations
instead of their explicit form. For these particular
integrals they were first considered
in \cite{Hoog}. One can get such relations from the identity:

\vskip 0.5cm
\begin{equation}
\int \frac{d^d~k_1~d^d~k_2}{[k_2^2-m_2^2]^{\beta} }
{}~~ \frac{\partial}{\partial k_{1 \mu}}
\left( \frac{A~ k_{1 \mu}-~B~k_{2\mu}}
{[k_1^2 -m_1^2]^{\alpha} [(k_1-k_2)^2+m_3^2]^
 {\gamma }} \right) \equiv 0
\end{equation}
%\vskip 1.0cm
with arbitrary constants $A$ and $B$.
The solution of recurrence relations for Feynman diagrams
was outlined in detail in \cite{CT}, \cite{Brod}. Exploiting the ideas
given there, one can get the recurrence
relations:
\vskip 0.3cm

\begin{eqnarray}
J_{\alpha \beta \gamma}~~~~&=&~~~~\sum J_{\alpha ' \beta ' \gamma '}\\
&&\nonumber \\
 \alpha+\beta+\gamma=R~~&&~~\alpha '+\beta '+\gamma ' \leq R-1,  \nonumber
\end{eqnarray}
%\vskip 0.3cm
where $R$ is  some integer number.
With the help of these relations,  $J_{\alpha \beta \gamma}$
with integer $\alpha, \beta, \gamma$
can be expressed in terms of trivial tadpole integrals (\ref{tadpole})
and one  "master" integral:

\vskip 0.5cm

\begin{equation}
J_{111}=\int \frac{d^d~k_1~d^d~k_2}{[k_1^2-m_1^2][k_2^2-m_2^2]
 [(k_1-k_2)^2-m_3^2]}
\end{equation}
with several permutations of the masses.
For arbitrary $d$ it can be represented as a combination
of hypergeometric  functions $~_2F_1$ \cite{Recur} and
its expansion for $\varepsilon \rightarrow 0 $
can be obtained from the  integral representation
of the latter.

Using this expansion and certain relations for dilogarithms,
  one  obtains \cite{Recur}:
\vskip 0.5cm
\begin{eqnarray}
\lefteqn{
J_{111}
= \pi^{4 - 2 \varepsilon} (m_3^2)^{1 - 2 \varepsilon}
\frac{A(\varepsilon)}{2}  }
\nonumber \\[0.3 cm] &&
\times
 \left \{ - \frac{1}{ \varepsilon^2} (1+x+y)
                + \frac{2}{\varepsilon} (x \ln x + y \ln y) \right.
\nonumber \\[0.3 cm] &&
\left.  -x \ln^2 x - y \ln^2 y
 + (1-x-y) \ln x \ln y - \lambda^2 \Phi (x,y)+ O(\varepsilon) \right\} \, ,
\end{eqnarray}
where
\begin{equation}
%\label{eq:Phi}
\Phi(x,y)=
\frac{1}{\lambda} \left (
2[\Sp {-\rho x} + \Sp { -\rho y } ] + \ln (\rho x) \ln(\rho y) +\ln
\frac{x}{y}~\ln \frac{1+\rho y}{1+ \rho x} + \frac{\pi^2}{3} \right ),
\end{equation}
$\Sp  { x }$  is Euler's dilogarithm  and
\begin{eqnarray*}
&& \lambda  \equiv \sqrt{ (1-x-y)^2 - 4 x y}, \;~~~~~
{}~~~~~~A(\varepsilon)=\frac{\Gamma^2(1+ \varepsilon)}
 {(1-\varepsilon)(1-2 \varepsilon) },\\
&& \\
&& \rho \equiv \frac {2}{1-x-y+\lambda}
\end{eqnarray*}
\vskip 0.5cm
For $\Phi $   we used representation given in \cite{UD}.
These results are valid for  $\lambda^2 \geq 0$
in the region \mbox{$ \sqrt{x}\, + \sqrt{y} \leq 1 $}. By permutation of $ m_1,
m_2, m_3$
one obtains results for other domains \cite{Recur}.

It is important to note that after partial fraction decomposition
the Taylor coefficients  of any two-loop
diagram
are expressible  in general as a combination
of $J_{111}$, $\ln x, \ln y $ and ratio of  polynomials
in (ratios of some squared masses)  $x$ and $y$, i.e.
this is not so only for two-point
functions.
  At the three-loop level the situation is more
complicated. For the time being there exits an algorithm \cite{Brod}
for the evaluation of the  Taylor coefficients
of a particular type
of three-loop  diagrams with only one mass different
from zero.
The generalization of this algorithm to similar types
 of diagrams with  one mass is rather
complicated, but possible. Any algorithm
for three-loop  diagrams with several different non zero
masses will be much more
elaborate.

\vglue 0.6cm
{\elevenbf\noindent 3. Expansion of three-point functions in terms of
external momenta squared}
\vglue 0.2cm
Here we have two independent external momenta in $d$ dimensions.
The general expansion of (any loop) scalar 3-point function with its
momentum space representation $C(p_1, p_2)$ can be written as
\begin{equation}
\label{eq:exptri}
C(p_1, p_2) = \sum^\infty_{l,m,n=0} a_{lmn} (p^2_1)^l (p^2_2)^m
(p_1 \cdot p_2)^n = \sum^\infty_{L=0} \sum_{l+m+n=L} a_{lmn}
(p^2_1)^l (p^2_2)^m (p_1 \cdot p_2)^n,
\end{equation}
where the coefficients 	$a_{lmn}$ are to be determined from the given diagram.
They are obtained by applying the differential operators
$\Box_{ij} = \frac{\partial}{\partial p_{i\mu}} \frac{\partial}{\partial
p_j^\mu}$
several times to both sides of (\ref{eq:exptri}):
\begin{eqnarray}
\begin{array}{llr}
\Box_{11} (p^2_1)^l (p_2^2)^m (p_1 \cdot p_2)^n & & \\
\\
= (p^2_2)^m \{ 2l (d + 2(l-1) +2n) \cdot (p^2_1)^{l-1} (p_1 \cdot p_2)^n
+ n(n-1) (p^2_1)^l p^2_2 (p_1 \cdot p_2)^{n-2}\}, & &
\end{array}
\end{eqnarray}
similarly for $\Box_{22}$ and
\begin{eqnarray}
\begin{array}{l}
\Box_{12} (p^2_1)^l (p_2^2)^m (p_1 \cdot p_2)^n\\
\\
= \{ n (2(l+m)+n+d-1) (p^2_1)^l (p^2_2)^m (p_1 \cdot p_2)^{n-1} +
4l\cdot m (p^2_1)^{l-1} (p^2_2)^{m-1} (p_1 \cdot p_2)^{n+1} \}.
\end{array}
\end{eqnarray}

This procedure results in a system of linear equations for the $a_{lmn}$. For
fixed $L$ (see equation (\ref{eq:exptri})) we obtain a system of $(L+1)(L+2)/2$
equations
of which, however, maximally $\left[ L/2 \right] + 1$ couple ($\left[
x\right]$ standing
here for the largest integer $\le x$). These linear equations are easily
solved with REDUCE \cite{REDUCE}, e.g., for arbitrary $d$.

For the purpose of demonstrating the method, we confine ourselves to the case
$p^2_1 = p^2_2 = 0$, which is e.g. physically realized in the case of the
Higgs decay into two photons ($H \to \gamma \gamma$) with $p_1$ and $p_2$
the momenta of the photons. In this case only the coefficients $a_{00n}$ are
needed. They are each obtained from a "maximally coupled" system of
$\left[ n/2\right]
+1$ linear equations. Solving these systems of equations we obtain a sequence
of differential operators ($Df$'s) which project out from the r.h.s. of
(\ref{eq:exptri})
the coefficients $a_{00n}$:
\begin{equation}
Df_{00n}=\sum^{[n/2]+1}_{i=1}\frac{(-4)^{1-i}\Gamma(d/2+n-i)\Gamma(d-1)}
 {2 \Gamma(i) \Gamma(n-2i+3) \Gamma(n+d-2) \Gamma(n+d/2)}(\Box_{12})^{n-2i+2}
 (\Box_{11}\Box_{22})^{i-1},
\end{equation}
where the sum of the exponents of the various $ \Box ' $ s is equal $n$.

The first few ($Df_{000} = 1$) for arbitrary $d$ are:

%
%     equations (4)
%
\begin{eqnarray}
\label{Df00n}
Df_{001} & = & \frac{1}{d} \Box_{12}, \qquad \qquad Df_{002} =
- \frac{1}{2(d-1)d(d+2)} \{ \Box_{11} \Box_{22} - d \Box^2_{12} \}, \nonumber\\
Df_{003} & = & - \frac{1}{2(d-1)d(d+2)(d+4)} \{ \Box_{11} \Box_{22}
\Box_{12}	- \frac{1}{3} (d+2) \Box^3_{12} \},\nonumber\\
Df_{004} & = & \frac{1}{8(d-1) d(d+1)(d+2)(d+4)(d+6)} \nonumber\\
& & \left\{ \Box^2_{11} \Box^2_{22}	- 2(d+2) \Box_{11} \Box_{22} \Box^2_{12} +
\frac{1}{3} (d+2)(d+4) \Box^4_{12} \right\} , \cdots .
\end{eqnarray}

Applying the operator $Df_{00n}$ to the (scalar) momentum space integral
$C(p_1,p_2)$ and putting the external momenta equal to zero, yields the
expansion coefficients $a_{00n}$.
Similarly we found for the coefficients $a_{l0n}$ the following
projection operators:
\begin{equation}
Df_{l0n}=\frac{\Gamma(\frac{d}{2}+n)} {\Gamma(l+1)
 \Gamma( \frac{d}{2}+l+n)} \left ( \frac{\Box_{11}}{4} \right )^l Df_{00n}
\end{equation}
For $n=0$ this reproduces the projection operators in (\ref{taylor2}).

Here we have not addressed the question of regularisation and
renormalisation of divergent diagrams. After subtraction
of divergent subgraphs only the first few terms in the Taylor expansion will be
 affected by ultraviolet divergences, which can then
easily eliminated by introducing counterterms. The
subtraction terms can be obtained in the same manner.

It is instructive to first look at the one-loop integral ($d=4$ since
the integral is finite):
\begin{equation}
C_0(m_1, m_2, m_3; p_1, p_2) = \frac{1}{i\pi^2} \int
\frac{d^4 k}{((k+p_1)^2 - m^2_1) ((k+p_2)^2 - m^2_2) (k^2 - m^2_3)}	.
\end{equation}

Performing the above procedure, it is seen that for $d=4$ the coefficients
$a_{00n}$ can be written as
\begin{equation}
\label{a00n}
i\pi^2 a_{00n} = \frac{2^n}{n+1} \int d^4 k \frac{(k^2)^n}{c^n_1 c^n_2}
\cdot \frac{1}{c_1 c_2 c_3}
\end{equation}
with $c_i = k^2 - m^2_i~ (i = 1, 2, 3)$. For $m_3 \not= 0$ we have
\begin{equation}
a_{000} = \frac{1}{i\pi^2} \int \frac{d^4 k}{c_1 ~ c_2 ~ c_3} =
C_0 (m_1, m_2, m_3; 0, 0) = \frac{1}{m^2_3} g(x,y),
\end{equation}
where $g(x,y) = \frac{1}{x-y}	\left( \frac{x \ln x}{1-x} -
\frac{y \ln y}{1-y}\right)$.
 The function $g(x,y)$ can now be considered as
{\it generating function} of the scalar one-loop vertex since all higher
negative powers of the $c_i's$ can be obtained by differentiation with respect
to $m^2_i$: writing $k^2/c_1 = 1 + m^2_1/c_1$, we finally have
\begin{equation}
\label{e22}
a_{00n} = \frac{2^n}{(n+1)!} \frac{1}{(m^2_3)^{n+1}}
\frac{\partial^n}{\partial y^n} \sum^n_{\nu=0} {n \choose \nu}
\frac{x^\nu}{\nu !} \frac{\partial^\nu}{\partial x^\nu} g(x,y),
\end{equation}
which can easily be evaluated with the help of a formulae manipulating
program.

Similarly we proceed in the two-loop case: starting e.g.~from the scalar
integral
\begin{eqnarray}
\label{treug2}
\begin{array}{l}
C(m_1, \cdots, m_6; p_1, p_2)\\
\\
= \frac{1}{(i\pi^2)^2} \int
\frac{d^4 k_1 d^4 k_2}{((k_1 + p_1)^2 -m^2_1)((k_1 + p_2)^2 - m^2_2)
((k_2 + p_1)^2 - m^2_3) ((k_2 + p_2)^2 - m^2_4) (k^2_2 - m^2_5)
((k_1 - k_2)^2 - m^2_6)}
\end{array}
\end{eqnarray}
in analogy to (\ref{a00n}) we now have with obvious abbreviations
\begin{equation}
\label{e24}
(i\pi^2)^2 a_{00n} = \frac{2^n}{n+1} \int d^4 k_1 d^4 k_2 F_n \cdot
\frac{1}{c_1~ c_2~ c_3 ~ c_4~ c_5 ~ c_6},
\end{equation}
where
\begin{equation}
\label{Fn}
F_n = \sum^n_{\nu=0} c_1^{-(n-\nu)} c_3^{-\nu} \sum^n_{\nu^\prime=0}
c_2^{-(n-\nu^\prime)} c_4^{-\nu^\prime} \cdot A^n_{\nu\nu^\prime} (k_1, k_2),
\end{equation}
and
\begin{equation}
\label{hernja}
A^n_{\nu\nu^\prime} (k_1, k_2) = \sum_{0\le 2\mu \le \nu + \nu^\prime \le
n+\mu} a^{n\mu}_{\nu\nu^\prime} (k^2_1)^{n-(\nu + \nu^\prime)+\mu}
(k^2_2)^\mu (k_1~ k_2)^{\nu + \nu^\prime - 2\mu},
\end{equation}
$a^{n\mu}_{\nu\nu^\prime}$ being rational numbers with the properties
\begin{equation}
\label{AAA1}
a^{n\mu}_{\nu\nu^\prime} = a^{n\mu}_{\nu^\prime\nu}	\qquad \mbox{and}
\qquad \sum_\mu a^{n\mu}_{\nu\nu^\prime} = 1,
\end{equation}
the first property being equivalent to $A^n_{\nu\nu^\prime} (k_1, k_2) =
A^n_{\nu^\prime \nu}(k_1, k_2).$ Besides that we also have the nontrivial
symmetry property
\begin{equation}
A^n_{\nu\nu^\prime} (k_1, k_2) = A^n_{n-\nu^\prime, n-\nu} (k_2, k_1)
\end{equation}
such that the essentially different $A^n_{\nu\nu^\prime}$ are the following
ones:
%\begin{equation}
\[
A^n_{\nu\nu^\prime}, \qquad \mbox{with} \quad	\nu \le \nu^\prime \qquad
\mbox{and} \qquad \nu + \nu^\prime	= 2, \cdots , n \quad (0 \le \mu \le
\left[ \frac{\nu + \nu^\prime}{2} \right], \quad \mbox{see (\ref{hernja}))}
\]
%\end{equation}
while we have explicitly $A^n_{00} = (k^2_1)^n$ and $A^n_{01} =
(k^2_1)^{n-1}	(k_1 \cdot k_2)$ or $a^{n0}_{00} = a^{n0}_{01} = 1$
(see (\ref{AAA1})). Finally we observe $a^{n0}_{02} = 4/3$ independent of $n$.
Some of the other higher coefficients are given explicitly
in  Appendix   A.

In the	two-loop case     we consider
as an example $m_6 = 0$ (gluon exchange) and all other masses $=m_t$
(top mass).
The "generating functions", i.e.~the diagrams with all
external momenta equal to zero, are now taken from Ref.~\cite{Bij/Ve}.
 In order to properly generate by differentiation the higher
negative powers of $c_1, c_2$ and $c_2, c_4$, respectively, we have to
introduce even in this case different masses corresponding to $c_i's$ with
different integration momenta and take the limit $m_1, m_2 \to m_t$ at the
end. In the notation of Ref.~\cite{Bij/Ve} we thus obtain
\begin{eqnarray}
\label{J231}
\int \frac{d^4 k_1 d^4 k_2}{(k^2_1 - m^2_1)^2 (k^2_2 - m^2_2)^3
(k_1 - k_2)^2} & = & < 2m_1 | 3m_2 |0>\nonumber\\
= \qquad \qquad \frac{1}{2} \frac{\partial^2}{\partial (m^2_2)^2} <2m_1|m_2|0>
& = &
(i\pi^2)^2 \frac{1}{2}	\frac{1}{(m^2_1)^2} \frac{\partial^2}{\partial a^2}
f(a,0)
\end{eqnarray}
with $f(a,0) = \Sp {1} - \Sp{a} - \ln a \cdot \ln (1-a)$ and $a=m^2_2/m^2_1$.
 After differentiation we
finally have as generating function for the diagram under consideration
\begin{equation}
g(a) = \frac{1}{2} \frac{1}{(m^2_1)^2} \left\{ \frac{1}{a} \frac{1}{1-a}
+ \frac{lna}{(1-a)^2} \right\} .
\end{equation}
 For arbitrary $d$, the integrals of the type (\ref{J231}) with arbitrary
powers of the scalar propagators can be
expressed in terms of hypergeometric functions:
\begin{eqnarray}
\label{eq:mast2m}
&&\int \frac{d^{d}k_1 d^{d}k_2}{\pi^d (k_1^2-m_1^2)^
 {\alpha}(k_2^2-m_2^2)^{\beta}(k_1-k_2)^{2\gamma}}= \nonumber\\
\nonumber\\
&&~~~~(-1)^{\alpha+\beta+\gamma}~\frac{\Gamma(\alpha+\beta+\gamma-d/2)
\Gamma(d/2-\gamma)
 \Gamma(\alpha+\gamma-d/2) \Gamma(\beta+\gamma-d/2)}
{\Gamma(\alpha) \Gamma(\beta) \Gamma(d/2)\Gamma(\alpha+\beta+2\gamma-d)
 (m_2^2)^{\alpha+\beta+\gamma-d}} \nonumber\\
\nonumber\\
&&~_2F_1(\alpha+\beta+\gamma-d,\alpha+\gamma-d/2,\alpha+\beta+2\gamma
 -d,z),
\end{eqnarray}
where $z=1-m^2_1/m^2_2$, which is also particularly useful for the equal
mass case, see Sect. 5.

For the two-loop case under consideration (see(\ref{treug2})), as can be
seen from the
above ((\ref{e24}) and following), it is more complicated
to write down
the analog of (\ref{e22}) in a compact form.
 In the general mass case, to calculate the coefficients
$a_{00n}$ in (\ref{e24}), one can proceed as follows:
in order to cancel powers of $k_1^2$ and $k_2^2$ (substituting
$k_1^2=c_i+m_i^2$ (i=1,2) and $k_2^2=c_i+m_i^2$ (i=3,4) )
as much as possible, one writes for the scalar product in (\ref{hernja}):
\begin{equation}
k_1 k_2=\frac12(k_1^2+k_2^2-m_6^2)-\frac12 c_6 \equiv k^2-\frac12 c_6
\end{equation}

and by stepwise reducing higher powers
( $ {\nu} + {\nu}^{\prime}-2{\mu} =  {\lambda} $ )
\begin{equation}
\label{kaka}
(k_1 k_2)^{\lambda}=(k^2)^{\lambda}+\left[(k^2)^{\lambda-1}+
 k_1 k_2 (k^2)^{\lambda-2}+ \dots + (k_1 k_2)^{\lambda-1}\right]
(-\frac{1}{2}c_6).
\end{equation}

In the second term $c_6$ cancels after insertion into (\ref{e24})
such that only factorizing one-loop contributions
are obtained from it. Moreover, in the square bracket of
(\ref{kaka}) only even powers of $k_1k_2$ contribute after integration.
The "genuine" two-loop contributions are then obtained
by replacing $k_1k_2$ in (\ref{hernja}) by $k^2 \equiv
\frac12(k_1^2+k_2^2-m^2_6)$
according to the first term in (\ref{kaka}).

  Introduction of ${c_i}^{\prime}$s in $k^2$ is not unique and
different choices for particular masses may be advantageous.
Nevertheless, simple power counting shows that in (\ref{Fn})
all numerator powers of $k_1^2$ as well as of $k_2^2$ can be cancelled
separately such that only "standard bubbles"  with no
integration momenta in the numerator remain.

 Even if compact formulae like (\ref{e22}) for the one-loop
three point functions cannot be obtained easily
in the general mass case of two loops, the above procedure
obviously can be the basis for a formulae manipulating
algorithm to calculate the higher order two-loop Taylor
coefficients (\ref{e24}). The central point of such an algorithm
is obviously the representation of the integrand in terms
of (\ref{Fn}) and (\ref{hernja}), which so far was obtained by inspection
of the results obtained as solution of the above mentioned
system of linear equations.
Moreover this representation also reflects the particular
topology of the diagram under consideration.
A similar procedure will, however, undoubtedly yield
corresponding formulae for other topologies in the general
mass case as well.

   Thus it is of interest to discuss the relevance of this
approach compared to the procedure predominantly
used at present to calculate the Taylor coefficients of
Feynman diagrams by means of recurrence relations.
These latter ones are obtained by the method of partial
integration, which is in general applicable for the arbitrary
mass case.
 Indeed, two-loop bubble diagrams can always be reduced by partial
fraction to integrals with at most 3 different masses and these
can be evaluated with the help of recurrence relations, which are
solvable in the sense that all integrals can be reduced by their
application to "master integrals" (\cite{bft}, \cite{FJT}).
 On the other hand, however, the above procedure
is more transparent and in any case is directly
applicable.

 For the equal mass case (with possibly an additional
zero mass) recurrence relations were quite successful \cite{Brod}.
 As seen from the above discussion, however, this case may
in some situations be less easy to handle by our new approach
since higher powers of scalar propagators are produced
by differentiation and for this purpose, as described above,
at least two different masses must be taken into account.
This may be a problem for three-loop calculations.
If, however, the three-loop general mass "master integrals"
can be found,  our approach may be simpler than the use of
recurrence relations since the possibility of their resolution
and "completeness"  is not yet fully understood.
In this sense the two approaches may well be complimentary
to each other and preference may be given to
them depending on the particular situation.\\

\vglue 0.6cm
{\elevenbf\noindent 4. The method of analytic continuation}
\vglue 0.4cm

Assume, the following Taylor expansion of a scalar diagram or a
particular amplitude is given:
\begin{equation}
C(p_1, p_2, \dots) = \sum^\infty_{m=0} a_m y^m \equiv f(y)
\label{orig}
\end{equation}
and the function on the r.h.s.~has a cut for $y \ge y_0$.
We have e.g. $y = \frac{q^2}{4m_e^2}$ for the photon propagator,
$y = \frac{q^2}{m_e^2}$ for the electron propagator function $I_3$,
and in the above case of $H \to \gamma\gamma$ one would introduce
$y =\frac{q^2}{4m_t^2}$ with $ q^2 = (p_1 - p_2)^2$ as
adequate variable. For these cases $y_0=1$.

 Our proposal for the evaluation
of the original series is in a first step a conformal mapping of the
cut plane into the unit circle and secondly the reexpansion
of the function under consideration
into a power series w.r.to the new conformal variable.
A variable often used \cite{Omega} is
\begin{equation}
\omega=\frac{b-\sqrt{1-\frac{y}{y_0} }}{b+\sqrt{1-\frac{y}{y_0}}}.
\label{omga}
\end{equation}
\renewcommand{\thefootnote}{\fnsymbol{footnote}}
Considering it as conformal transformation,
the y-plane, cut from $y_0$ to $+ \infty$, is mapped into the unit
circle \footnote{"Was au{\ss}erhalb dieses Kreises ist, ist das
Nichts. Was innerhalb dieses Kreises ist, ist das All",
Klabund,  Der Kreidekreis.  }
and the cut itself is mapped on its boundary, the upper
semicircle corresponding to the upper side of the cut.
The origin goes into the point
$(b-1)/(b+1)$. By suitable choice of $b$ it can be placed
on the real $\omega$ axis anywhere between $-1$ and $+1$.
For our applications we shall take $b=1$, i.e. the origin
of the expansion will be at $\omega=0$.\\
%%%%%%%%%%%%

After conformal transformation it is suggestive to improve the
convergence
of the new series w.r.to $\omega$ by applying one of the numerous
summation methods \cite{Sha},\cite{Weniger} most suitable for our problem.
We obtained the best results with the Pad\'e method and partially
also with the Levin $v$ transformation. For $b=1$ the expansion of
$f(y)$ in terms of $\omega$ is:

\begin{equation}
f(y(\omega))=\sum_{s=0}^{\infty} \omega^{s} \phi_s,
\label{foty}
\end{equation}
where
\begin{eqnarray}
\phi_0&=&a_0 \nonumber \\
&&\nonumber \\
\phi_s&=&\sum_{n=1}^{s}a_n(4y_0)^n\frac{\Gamma(s+n)(-1)^{s-n}}
 {\Gamma(2n) \Gamma(s-n+1)},~~~~~s \geq 1.
\end{eqnarray}

Eq.(\ref{foty}) will be used for the analytic continuation of $f$
into the region of analyticity ($y<y_0$; observe that the series
(\ref{orig}) converges for $\left| y \right|<y_0$ only)
and in particular for the continuation on the cut ($y>y_0$).
In this latter case we write
\begin{equation}
\omega=exp[i \xi(y)],~~~~~~~~~~~{\rm with}~~~ \cos \xi=-1+2~\frac{y_0}{y}
\end{equation}
and hence

\begin{equation}
f(y(\omega))=a_0+\sum_{n=1}^{\infty}\phi_n \exp i n \xi(y)
\label{foncut}
\end{equation}

In any case we have $\left| \omega \right| \leq$ 1 and we will show in the
following how to sum the above series.

Pad\'e approximations are indeed particularly well suited for the summation
of the series under consideration. In the case of two-point functions they
could be shown in several cases (see e.g. \cite{bft} ) to be of
Stieltjes type (i.e. the spectral density is positive). Under this
condition the Pad\'e's of the original series (\ref{orig}) are guaranteed
to converge in the region of analyticity. For the three-point function
under consideration ($H \to \gamma\gamma$), however, the obtained result
shows that the series is not of Stieltjes type (i.e. the
imaginary part changes sign on the cut).

   Having performed the above
 $ \omega $ transformation (\ref{omga}), however, it is rather the
Baker-Gammel-Wills {\it conjecture } ( see \cite{BGW} ), which applies
and which we quote in the following in its original form:

{\bf Conjecture} : If P(z) is a power series representing a function
which is regular for $\left| z \right| \leq$ 1, except for m poles
within this circle and except for z=+1, at which point the function
is assumed continuous when only points $\left| z \right| \leq$ 1 are
considered, then at least a subsequence of the $\left[ M/M \right]$
Pad\'e approximants converges uniformly to the function (as M tends
to infinity) in the domain formed by removing the interiors of small
circles with centers at these poles.

So far, there are no known counter examples nor any valid proof of it.
In the present case of Feynman diagrams, these are analytic in the whole
cut plane, i.e. after $ \omega $ transformation within the whole circle
of radius 1 in $ \omega $. There can be in general e.g. logarithmic
(or even worse)
singularities at threshold (y=$ \omega $ =1), but in a "refinement"
of the present procedure these may be even factored out.
In the latter case one exactly satisfies the assumptions of the BGW
conjecture, while in general one may be confident to obtain even without
refinement excellent results as is demonstrated in the following
applications. This is not unexpected since the above conjecture does
not claim to fully exhaust the convergence properties of Pad\'e
approximants.

 A convenient technique for the evaluation of Pad\'e approximants
is the $\varepsilon$-algorithm
of~\cite{Sha}. In general, given a sequence
$\{S_n|n=0,1,2,\ldots\}$, one constructs a table of approximants using
\begin{equation}
T(m,n)=T(m-2,n+1)+1/\left\{T(m-1,n+1)-T(m-1,n)\right\},
\label{eps}
\end{equation}
with $T(0,n)\equiv S_n$ and $T(-1,n)\equiv 0$. If the sequence $\{S_n\}$ is
obtained by successive truncation of a Taylor series, the approximant
$T(2k,j)$ is identical to the $[k+j/k]$ Pad\'{e} approximant~\cite{Sha},
derived from the first $2k+j+1$ terms in the Taylor series.

A slight generalization of the above also applies to the multivariate
case of the three-point function (\ref{eq:exptri}) as well
(for $H \to \gamma\gamma$   we have
$p_1^2=p_2^2=0$ such that only a single variable expansion has to be
investigated). In general we propose to proceed as follows: at first,
for physical reasons, an expansion in terms of   squares of the external
momenta seems adequate
(writing e.g. $p_1 \cdot p_2 = \frac{1}{2}(p_1^2+p_2^2-q^2), q=p_1-p_2)$.
Then in each variable the $ \omega $ transformation is supposed to be
performed separately, yielding a series of the form

\begin{equation}
\sum^\infty_{l,m,n=0} b_{lmn} {\omega}_1^l {\omega}_2^m
{\omega}_3^n = \sum^\infty_{L=0} \sum_{l+m+n=L} b_{lmn}
{\omega}_1^l {\omega}_2^m {\omega}_3^n,
\label{oma}
\end{equation}
in analogy to (\ref{eq:exptri}). The final question is how to
perform the summation in
this case. As we have seen above, the $\varepsilon$-algorithm only
supposes a sequence of elements $\{S_n|n=0,1,2,\ldots\}$. Thus
it appears natural to take as appropriate sequence the partial sums
in (\ref{oma}) over L and performing the $\varepsilon$-algorithm with these
as input. In this manner we can in principle apply the method to n-point
functions with any n (n=2,3,4, \ldots = self- energies, vertices, boxes etc.
).
Another possibility would be  the application of direct generalisations
of Pad\'e approximants to the multivariable case, like Chisholm
approximants \cite{Chisholm}. For  a general introduction to these
methods see also Baker et al. in \cite{Sha}, vol.14.
Numerical examples have so far not been calculated in the more general
cases.

   In special situations, instead of using Pad\'e's, the
Levin $v$ -transformation yields even better results. It is defined by

\begin{eqnarray}
\label{levin}
T(n,k)&=&\frac{\sum_{m=0}^{k}(S_{n+m}(\rho_{n+m}-1)/a_{n+m+1})
 (n+m+1)^{k-1}(-1)^m (^k_m)}
 {\sum_{m=0}^{k}((\rho_{n+m}-1)/a_{n+m+1})(n+m+1)^{k-1}(-1)^m
 (^k_m)},  \\
&& \nonumber \\
a_n&=&S_n-S_{n-1},\nonumber  \\
&& \nonumber \\
\rho_n&=&a_{n+1}/a_n. \nonumber
\end{eqnarray}

We stress that it is an absolute necessity for the above procedures
to work successfully, that the Taylor coefficients are known with very
high precision. Only then will the delicate method of analytic continuation
really work (see Sect.5). This is, however,
no limitation if one uses modern formulae manipulating systems, e.g.	In
particular we have performed for this reason all the calculations
(including the small part of the "numerics")
in REDUCE, which allows (among other languages) for
arbitrary length precision
of real (and rational) numbers.
It may even
turn out that adding Feynman diagrams is best performed on the level of these
coefficients (if not already on the level of the generating functions
which were introduced in Section 3).

As a further remark we want to stress the following general property of the
proposed method: may it take enormous effort to calculate the expansion
coefficients of the original series, they can, however, be calculated
analytically once for ever as a function of the mass ratios and following
the demonstrated procedure, the diagrams can then be obtained in a very
simple manner in the whole kinematical region of interest. Beyond that they
may finally use relatively little computer time. Even in the one-loop
case this is of great relevance in the comparison with experiments
(LEP e.g), where radiative correction codes often enter MC programs.

Finally we mention, as will be demonstrated in the next section, that the
above mapping ($ \omega $ transformation) in combination with Pad\'e
approximants ($ \omega $P) is particularly effective for the problems under
consideration. It is, however, always worth to investigate other mappings
as well. Here this additional procedure serves as a precision test for
the obtained results in the case of the three-point function
$H \to \gamma\gamma$. This is important since the MC method of Ref.
\cite {Fuji} does by far not reach the precision our method provides.
Following the lines of Ref. \cite{Fle}, one can choose

\begin{equation}
\label{e43}
x(y) = \frac{1}{2} ln \frac{1-y}{1+y},~~~~~y(x)=-\frac{e^x-e^{-x}}
 {e^x+e^{-x}},
\label{map}
\end{equation}
i.e.~a mapping of the whole $y$-plane on the strip with $-\pi/2 \le Im(x)
\le \pi/2$ in the $x-$plane. In particular the region of convergence of the
series (\ref{orig}), i.e.~$1 \ge y \ge -1$ is mapped on the real $x$-axis
 $- \infty
\le x \le + \infty$ with $x(0) = 0$. The (upper side of the) cut $(1 \le y
\le + \infty)$ is mapped into $-\infty \le Re(x) \le 0$ and $Im(x) = - \pi/2$.
Furthermore the mapping (\ref{map}) has the nice property
\begin{equation}
\label{e44}
Re(x(z > 1)) = x ( \frac{1}{z} < 1) < 0.
\end{equation}

Thus we obtain on the cut an exact representation for the function $f(y)$ by
expanding in a Taylor series in $x$ around $x_C \equiv x(1/z)$, e.g.
Using $y^{\prime}(x)=y^2-1$ we finally	have
\begin{equation}
\label{e45}
f(z > 1) = \sum^\infty_{n=0} \frac{\left\{ (y^2 -1) \frac{d}{dy} \right\}^n
f(y)\mid_{y = 1/z < 1}}{n!} (-i \frac{\pi}{2})^n ,
\label{series}
\end{equation}
which is convergent when no singular point of $f(y(x))$ lies inside the circle
with center $x_C$ and radius $\pi/2$. In particular all coefficients
of the series (\ref{series}) are convergent if the expansion (\ref{orig}) for
$f(y)$ is used
since the derivatives are taken at points $0 \le y = \frac{1}{z} =
\frac{4m^2_t}{q^2} < 1$ if $z > 1$ on the cut, i.e.~they are taken
in the region of analyticity of the original series. The term $n=0$ in
(\ref{series}) is just the original series, but now taken for $y = 1/z < 1$.
What concerns the higher terms in (\ref{series}), it turns out to
be of great use near threshold to factor out an overall $(y^2 -1)$ and apply
Pad\'e approximations to the remaining coefficient-series (sub-series).
Observe: if in (\ref{orig}) the	coefficients $a_0, \cdots, a_M$ are known,
then in the sub-series of
(\ref{series}) only coefficients $a_0^{(n)}, \cdots, a^{(n)}_{M-n}$ are
completely known and only these should be taken into account.\\

\vglue 0.6cm
{\elevenbf\noindent 5. Applications}
\vglue 0.4cm

 We applied our method to the evaluation of   one-
and two- loop integrals, which we consider as representative for
applications in QED, QCD and electroweak processes
in general. We present
 results for the two-loop photon vacuum polarisation
function, the
two- loop fermion "master integral" $I_3$ \cite{DB}
and the two-loop three point scalar integral with the
kinematics of the decay $H \rightarrow \gamma \gamma$.

 a) Vacuum polarization function.

 In the on-shell renormalization scheme of conventional
 QED, the renormalized photon propagator has the dominator
$(1+\Pi)$, where the vacuum polarisation function,
$\Pi$, vanishes at $q^2=0$. Up to two-loops it can be written

\begin{equation}
 \Pi(z)=\left ( \frac{\alpha}{4\pi}\right ) \Pi_1(z)+ \left (
 \frac{\alpha}{4\pi} \right) ^2 \Pi_2(z) + O(\alpha ^3),
\end{equation}
where $ z \equiv q^2/4m^2$.
 The analytic expression for $\Pi_2(z)$ was given in \cite{photon2}
 and for  arbitrary space-time dimension
 in \cite{bft}.

  Tables 1 and 2 contain the results of our calculation with
$\omega P$ for space-like and time-like $q^2$, respectively .
Here $E_4(N)=|\Pi_2^{[N/N]}/\Pi_2^{exact}-1|$ is the relative
error of the $[N/N]$ Pad\'e approximation, the index 4 referring
to $d=4$. As is seen from the tables, the Pad\'e approximants
are fast and reliably converging even in the time-like region
on the cut. Of course, in the latter case they supply
simultaneously real and imaginary part and the real part
has similarly good convergence properties, which we
don't show, however.

In many applications one doesn't need very
high precision of the Feynman diagrams,
in particular in the two-loop case.
 Therefore, given a certain requested precision,
one needs only a limited number of
coefficients  even though this number will in general
depend on the value of the kinematical variables under
consideration.
Thus it is interesting to see how many
coefficients are needed in order to achieve at least $1\%$
accuracy, where the scale for $q^2$ is set by its
threshold value $q^2_{thr}$ and the typical
domain of interest is $q^2_{thr} \leq  q^2 \leq 100 q^2_{thr}$.
In the present case of the photon self-energy, the threshold
is  $q^2_{thr}=4m^2$ such that
we are mainly interested in values $|q^2/m^2| \leq 400$.
{}From Table 1 it is seen that 8 coefficients
are sufficient for the $1\%$ accuracy in this domain of the spacelike
region. Increasing $q^2$ by one order of magnitude,
only six more coefficients are  enough  for the same precision.
In the timelike region on the cut, 14 coefficients
suffice and also here eight more coefficients are enough
for $|q^2/m^2| \leq 4000$. Moreover, as the tables demonstrate,
for smaller $q^2$ ' s the convergence is orders of magnitude
better, except very close to the threshold (4.01 in the Table 2)
where a singularity appears.

b) Two-loop fermion self- energy "master integral".

Similarly, good results were obtained for the two-loop
master integral $I_3$ \cite{DB},\cite{bft} of the electron
propagator:
\begin{equation}
\label{I3}
I_3\equiv \frac{-q^2}{\pi ^d} \int
\frac{d^d~k_1~d^d~k_2}{k_1^2[k_2^2-m^2][(k_1-q)^2-m^2](k_2-q)^2
 [(k_1-k_2)^2-m^2]}.
\end{equation}

For $d=4$ the threshold  value of this integral
is finite \cite{DB}:
\begin{equation}
\label{I3num}
I_3|_{q^2=m^2}=\frac32 \zeta(3) - \pi^2 \ln 2=-5.038 ~003~109~117~725...
\end{equation}

Table 3 demonstrates the convergence properties of $\omega P$
in this case.
 The integral $I_3$ is one of the most complicated
ones since it has two thresholds: one at $q^2=m^2$
(where according to (\ref{I3num}) $Im I_3=0$) and the next at $q^2=9m^2$.
Comparing the errors  with those of  Table 2 for the
photon self-energy  at the same scale, we see that below the second
threshold the convergence properties are similar.
Crossing, however, the second threshold, the convergence
properties are not that excellent anymore:
compare $E_4(10)$ at $q^2/m^2=40$ in Table 2 with $E_4(10)$
at $q^2/m^2=10$ in Table 3. Nevertheless,
taking into account more coefficients, we observe convergence even for
$q^2/m^2=100$ also in this case.

 As we have mentioned before, the convergence
properties of the real part are very similar to that
of the imaginary part and this is indeed so for the cases
shown in Tables 2 and 3.
The fermion self- energy integral $I_3$ at threshold,
however, is real and finite and unfortunately
the convergence properties of $\omega P$ are worse
in this case. For $q^2=m^2$ we have, e.g.,
$E_4(6)=4.5 \times 10^{-3}$, $E_4(10)=6.9 \times 10^{-4}$,
$E_4(14)=4.2 \times 10^{-4}$, $E_4(20)=1.1 \times 10^{-4}$,
and $E_4(40)=1.2 \times 10^{-5}$ (the diagonal
Pad\'e 's without $\omega$ transformation being
much worse).

In this case, however,  the Levin $v$ transformation
$T(1,k)$ (see \ref{levin}) improves the situation somewhat.
To obtain the same precision as above, the following
number of coefficients is needed, correspondingly:
8,~9,~12,~20,~54 (remember that for $E_4(N)$ $2N$ coefficients
are necessary) and with 80 coefficients the relative precision
is $5.2\times10^{-6}$.
This approach may become relevant for the calculation
of the magnetic moment of the electron in higher
than two-loop order.

c) Decay of $H \rightarrow \gamma \gamma $.
\vskip 0.2cm
 At the one loop level we investigated the integral

\begin{equation}
I_1=m^2 \int \frac{ d^d k_1}{i \pi^{d/2}}
\frac{1}{[(k_1+p_1)^2-m^2][(k_1+p_2)^2-m^2]
 [k_1^2-m^2]}.
\end{equation}

For $p_1^2=p_2^2=0$ and $d=4$ the integral can simply be
written as

\begin{equation}
I_1= \frac{\arcsin^2 \sqrt{z} }{2z}.
\end{equation}

At  threshold, i.e. $z=(p_1-p_2)^2/4/m^2=1$, $I_1$ is finite. With 21
coefficients $\omega P_{[10/10]}$ has a relative error of $1\times 10^{-16}$
while the Levin $v$ transformation is accurate up to a relative
error of $2\times 10^{-22}$. Apart from the fact that the achieved
precision is quite spectacular, again this result
shows that the Levin method gives better results at threshold.

In the two-loop case we study the integral (\ref{treug2})  with $m_6=0$ and
all other masses $m_i=m_t (i=1,..,5$). As explained above, using
(30), all Taylor coefficients can be expressed in terms of
$\Gamma$ -functions and a list of the first coefficients $a_{00n}$
($\equiv a_n; n=0,...,28)$ is given in the Appendix B.

The results for this case are given in Tables 4 and 5,
again below threshold and above (on the cut), respectively.
For comparison we use the values supplied to us by the
authors of Ref. \cite{Fuji}, who use for
the evaluation of the integral under consideration
the Monte Carlo (MC) method.

Table 4 shows, below threshold, perfect agreement
with the MC method within the statistical
errors and it is seen that our $\omega P$ method with
the given number of coefficients yields excellently
converging approximants. If we believe
in the precision of the $\omega P$'s to such
a degree as we find agreement with nearby approximants,
then the convergence is indeed spectacular and the obtained
precision goes far beyond the MC method. This is so, even
if we raise doubts about the last
decimal of the various entries in  Table 4.

Similarly, high precision is obtained on the cut, as
is demonstrated in Table 5. Here, both real and imaginary
part of the scalar two-loop $H \rightarrow \gamma \gamma$ integral are
shown, again in  comparison with the MC
method of Ref.\cite{Fuji}. As before, we consider the
domain $q^2_{thr} <q^2 \leq 100 q^2_{thr}$, where
$q^2_{thr}=4m^2_t ~(m_t=150 GeV)$. For $q^2$ close to the
threshold, the integral has a logarithmic singularity, but
still without "refinement", we obtain good stability of the approximants,
which improves to 8-10 decimals up to $q^2=10q^2_{thr}$
and even for $q^2=100q^2_{thr}$ is still excellent.
In the latter case the MC program of Ref. \cite{Fuji} was found to
provide unstable results for the real part.
Apart from that, there occur slight systematic deviations
between the two approaches, which are very likely, however,
solved in favor of our approach due to the high stability of the
Pad\'e approximants.

 This fact is even confirmed by performing an independent
calculation according to (\ref{series}). A slight
improvement of the basic formula (\ref{series})
can  still be obtained in the following manner:
we choose (see also the discussion after (\ref{e44})) $x_C \equiv x(\frac1z)
-i \frac{\pi}{4}r$ ($0 \leq r < 2$) and expand into a Taylor series
in $x$ around the line $Im (x)=-\frac{\pi}{4}r$ instead
of $Im (x) =0$ (r=0). From (\ref{e43}) we obtain $y_C=y(x_C)$
and instead of (\ref{e45}) we now have

\begin{equation}
\label{Joch_conf}
f(z > 1) = \sum^\infty_{n=0} \frac{\left\{ (y^2 -1) \frac{d}{dy} \right\}^n
f(y)\mid_{y = y_C}}{n!} (-i \frac{\pi}{4}(2-r))^n .
\end{equation}

For r=1, $y_C=e^{i\phi}$ and $\phi =2~ arctg ~z
-\frac{\pi}{2}$~($0 \leq \phi \leq \pi / 2$ for $1 \leq z \leq \infty $)
and the "subseries" are Fourier series as in the case of $\omega P$;
for $r>1$ we have $|y_C| > 1$ which is outside their domain of convergence.
Even for $r> 1$, however, the Pad\'e approximants can be used for the
summation of the subseries. Since they have, as explained
above, less and less completely known coefficients for the higher
derivatives, we observe that from a certain summation index on, the
Pad\'e approximants become quickly unstable; thus we sum (\ref{Joch_conf})
up to
such maximal $n=n_{max}$ for which they remain stable. In this manner
we arrive,e.g., at the following result for $q^2/{m_t}^2=10$ with
28 originally given coefficients (where also for the summation of the
main series in (51) Pad\'e approximants have been used):

$$Re(f)=-0.75692 ~~~~  and ~~~~Im(f)=-0.06138~~~; r=1, n_{max}=12 $$
$$Re(f)=-0.7569425~~~~ and ~~~~Im(f)=-0.061545~~; r=\frac{3}{2}, n_{max}=9 $$
which by inspection of Table 5, gives preference to our
value obtained with $\omega P$ in comparison with the
value obtained in Ref. \cite{Fuji}.

 While $\omega P$ is to be considered a very effective
algorithm, (\ref{e45})  and (\ref{Joch_conf}), respectively, are
representations of amplitudes on the cut, which may be rather
of analytic interest in general.

 As was mentioned already before, for our methods of analytic
 continuation to work with such high precision,
also the Taylor coefficients must be known with high accuracy.
In general we should know them analytically and then
approximate them with the desired precision.
A good example are the coefficients for the $H \rightarrow \gamma
\gamma$ decay, which can be represented as rational
numbers (see Appendix B) and which for our purpose were approximated with
a precision of 45 decimals using REDUCE.

 To demonstrate the relevance of high precision, we
repeated for the photon propagator the calculation in the
timelike region, keeping only 15 decimals, as are usually
available in standard FORTRAN compilers. Some relative
errors for the imaginary part of the photon propagator are
given for this case in Table 6. While for the low order
Pad\'e approximants ($[2/2]$  and $[4/4]$) the errors
are the same, small deviations show up for the $[6/6]$,
and for the higher approximants
several orders of magnitude are quickly lost as seen from
the table. Also the decrease of the errors in general
slows down considerably with increasing order.

The numerical instability, inherent to our nonlinear computational
schemes, was in the present paper eliminated by using REDUCE's
exact rational and floating-point arithmetics with a predefinable
degree of precision, and it is clear that in general
it is the easiest way to   use symbolic manipulation
programs  for the
realization of the $\epsilon$ algorithm or the Levin
transformation.   Once, however, speed is crucial, e.g.
when our methods are used in MC programs for comparison
with experimental data, it may become necessary to use special
properties  of the language C or to write even assembler routines
for the arbitrary length arithmetic.

%%%%%%%%%%%%%%%%%%88888

\vskip 1.0cm

\vglue 0.6cm
{\elevenbf\noindent 6. Conclusions}
\vglue 0.4cm

We have demonstrated, that the method of expanding Feynman
diagrams with respect to their external momenta squared
can be made a very effective tool to evaluate any amplitude
in the whole kinematical domain of interest. The method is quite
generally applicable to higher $n$- point functions,
higher loops, several nonvanishing masses, "tensor integrals",
etc.
In spite of that it is systematic and simple in comparison
with other methods, the most important simplification being that
the hard analytic work needs to be performed only for all external
momenta equal  zero. That this is possible, relates to the
miraculously precise procedure of combining a mapping,
mainly the "$\omega$- mapping", in combination with Pad\'e
approximants, which almost certainly will converge in
general, in the univariable case due to the Baker-Gammel- Wills
conjecture.
Here, for the purpose of analytic continuation, it is of great
importance to use symbolic manipulation, in particular to obtain
the needed Taylor coefficients to high precision. This is
possible  by means of formulae manipulating systems,
which in particular allow for arbitrary length precision.
For this purpose
we used  REDUCE \cite{REDUCE} and FORM \cite{FORM}.

 Knowing, finally, the Taylor coefficients of a certain
diagram, one has obtained an ever lasting piece of information
from which it is a relative small task to obtain the
corresponding Feynman diagram in the whole kinematical area
of interest. For this reason our present work may initiate a
new approach to practical calculations in perturbation
theory of quantum field theory even though, admittedly, quite
some problems have still to be worked out in greater detail,
like the convergence in the multivariable case, "generating
functions"  for three- loop diagrams with arbitrary masses,
diagrams with thresholds starting at zero momentum squared etc.
The most pressing  task for the near future will be the development of
algorithms for the fast calculation of the Taylor coefficients.

\bigskip
\noindent
{\elevenbf \noindent Acknowledgments}
\vglue 0.2cm
We are  grateful to J. Fujimoto for providing the numerical values
of Ref.~\cite{Fuji} for	comparison. We are also grateful to
D.Broadhurst and A.Davydychev for reading the text of manuscript
and  useful discussions.

\newpage

{\bf Appendix A.}

Coefficients $a_{\nu \nu'}^{n \mu}$ for various $n$.
$\nu+\nu'=2,...,n$ and $0 \leq \mu \leq [\frac{\nu+\nu '}{2}]$.
Since we have $\sum_{\mu} a_{\nu \nu'}^{n \mu}=1$ (see (27)) the
 coefficient with the highest $\mu$ is not given explicitly.
The coefficients are ordered with respect to increasing $\nu + \nu '$,
beginning with the lowest $\nu$.
\[
\begin{array}{lllllll}
n = 2 : & a^{20}_{11} = 1/3. & & & \\
n = 3 : & a^{30}_{11} = 4/9; & a^{30}_{03} = 2; & a^{30}_{12} = 2/9 .& & &\\
n = 4 : & a^{40}_{11} = 1/2; & a^{40}_{03} = 2; & a^{40}_{12} = 1/3; & & &\\
        & a^{40}_{04} = 16/5,& a^{41}_{04} = -12/5; & a^{40}_{13} = 1/5, & &
&\\
      & a^{41}_{13} = 11/10;& a^{40}_{22} = 4/45, & a^{41}_{22} = 22/45; & &
&\\
n =5 :& a^{50}_{11} =  8/15;& a^{50}_{03} = 2   ; & a^{50}_{12} =  2/5 ; & &
&\\
      & a^{50}_{04} = 16/5 ,& a^{51}_{04} =-12/5; & a^{50}_{13} =  8/25, & &
&\\
      & a^{51}_{13} = 24/25;& a^{50}_{22} = 4/25, & a^{51}_{22} = 12/25; & &
&\\
      & a^{50}_{05} = 16/3 ,& a^{51}_{05} =-16/3; & a^{50}_{14} = 16/75, & &
&\\
      & a^{51}_{14} =128/75;& a^{50}_{23} =  4/75,& a^{51}_{23} = 32/75; & &
&\\
n =6 :& a^{60}_{11} =  5/9 ;& a^{60}_{03} = 2   ; & a^{60}_{12} =  4/9 ; & &
&\\
      & a^{60}_{04} = 16/5 ,& a^{61}_{04} =-12/5; & a^{60}_{13} =  2/5 , & &
&\\
      & a^{61}_{13} = 13/15;& a^{60}_{22} =16/75, & a^{61}_{22} =104/225;& &
&\\
      & a^{60}_{05} = 16/3 ,& a^{61}_{05} =-16/3; & a^{60}_{14} = 16/45, & &
&\\
      & a^{61}_{14} = 68/45;& a^{60}_{23} =  8/75,& a^{61}_{23} = 34/75; & &
&\\
      & a^{60}_{06} = 64/7 ,& a^{61}_{06} =-80/7, & a^{62}_{06} = 24/7 ; & &
&\\
      & a^{60}_{15} = 16/63,& a^{61}_{15} =176/63,& a^{62}_{15} =-47/21; & &
&\\
      & a^{60}_{24}=64/1575,& a^{61}_{24}=704/1575,& a^{62}_{24}=16/21 ; & &
&\\
      & a^{60}_{33} = 4/175,& a^{61}_{33} =44/175,& a^{62}_{33} =3/7   ; & &
&\\
n =7 :& a^{70}_{11} =  4/7 ;& a^{70}_{03} = 2   ; & a^{70}_{12} = 10/21; & &
&\\
      & a^{70}_{04} = 16/5 ,& a^{71}_{04} =-12/5; & a^{70}_{13} = 16/35, & &
&\\
      & a^{71}_{13} =  4/5 ;& a^{70}_{22} =16/63, & a^{71}_{22} =  4/9 ; & &
&\\
      & a^{70}_{05} = 16/3 ,& a^{71}_{05} =-16/3; & a^{70}_{14} = 16/35, & &
&\\
      & a^{71}_{14} = 48/35;& a^{70}_{23} = 16/105,& a^{71}_{23}= 16/35; & &
&\\
      & a^{70}_{06} = 64/7 ,& a^{71}_{06} =-80/7, & a^{72}_{06} = 24/7 ; & &
&\\
      & a^{70}_{15}=64/147,& a^{71}_{15}=368/147,& a^{72}_{15} =-104/49; & &
&\\
      & a^{70}_{24}=64/735,& a^{71}_{24}=368/735,& a^{72}_{24} =472/735; & &
&\\
      &a^{70}_{33}=64/1225,& a^{71}_{33}=368/1225,& a^{72}_{33}=472/1225;& &
&\\
      & a^{70}_{07} = 16  ,& a^{71}_{07} =-24   , & a^{72}_{07} = 10   ; & &
&\\
      & a^{70}_{16} = 16/49,& a^{71}_{16}=232/49,& a^{72}_{16} =-246/49; & &
&\\
      & a^{70}_{25}=16/441,& a^{71}_{25}=232/441,& a^{72}_{25} =538/441; & &
&\\
      &a^{70}_{34}=16/1225,& a^{71}_{34}=232/1225,& a^{72}_{34}=538/1225
        \quad \mbox{etc.}
\end{array}
\]

\newpage
{\bf Appendix B.}
Taylor coefficient $a_{00n} \equiv a_n$ for the scalar integral
with the kinematics of the $H \rightarrow \gamma \gamma $ decay
according to  (23) and (34). An overall factor
$(\frac{1}{16 \pi^2})^2 \frac{1}{m_t^4}$ ($[GeV]^{-4}$) is taken out.
\[
\begin{array}{ll}
a_{0}=\frac{1}{4} & a_{1}=\frac{17}{144}\\
a_{2}=\frac{21}{400} &a_{3}=\frac{14857}{635040} \\
%\end{array}
%\]
a_{4}=\frac{55717}{5292000}&  a_{5}=\frac{1969949}{411642000}\\
a_{6}=\frac{112205382323}{51132111030000}&a_{7}=\frac{51839745827}{51132111030000}\\
a_{8}=\frac{774014944337}{1641908898630000}&
a_{9}=\frac{554222437367}{2514608355659400}\\
a_{10}=\frac{18901210440613}{182560566620872440}&
a_{11}=\frac{34011823443536821}{696451007757991736250}\\
a_{12}=\frac{73469051403634734553}{3177905948399716292508750}&
a_{13}=\frac{2492210424397715213}{226993282028551163750625}\\
a_{14}=\frac{554558587115114132753}{106056305658895293730153125}&
a_{15}=\frac{493942604968420733709634}{197844918903561555886138003125}\\
a_{16}=\frac{2390970211723833244102235042}{2001201354709525137788285901609375}&
a_{17}=\frac{181038174524415849605612756}{315979161269925021756045142359375}\\
a_{18}=\frac{1676442262041467600477129419999676}{6090230067169886628004299236651090015625}&\\
a_{19}=\frac{32272719643196898524719751009456}{243609202686795465120171969466043600625}&\\
a_{20}=\frac{4807871353613440589160177061808}{75290875108752192841884368573711949375}&
\end{array}
\]

\[
\begin{array}{ll}
a_{21}=\frac{4034419246662421906160632413314528}{130880496186718164652813432425466404103125}&\\
a_{22}=\frac{7221193650976946979268781960136850976
       }{484650477379417363709368140271502094393871875}&\\
a_{23}=\frac{18013224509615331344552281842106547013376}{
  2498050110572643565012653184339412295204146934375}& \\
a_{24}=\frac{15578909526611620403306533456511933062343936}{
    4459019447372168763547585934045850946939402277859375}& \\
a_{25}=\frac{581251584252075210856954583040719209869824}{
  343001495951705289503660456465065457456877098296875}&\\
a_{26}=\frac{230685031319614889642002712947387308039884288}{
 280381037656395284136867842442170835184128186779765625}&\\
a_{27}=\frac{10796905761405839554732050758878419622842640384}{
 27003594419803863054974892549689073885147242264685703125}&\\
a_{28}=\frac{4416531481278514440382313870488956528672122558464}{
  22710022907055048829233884634288511137408830744600676328125}
\end{array}
\]

\bigskip
\bigskip
\vglue 0.5cm

\pagebreak

\input results.tex

\end{document}